\documentclass[a4paper,twocolumn]{article}
\usepackage[latin1]{inputenc}
\usepackage{amsmath,amssymb}
\usepackage{theorem}
\usepackage{graphicx}
\usepackage{color}

\usepackage[bf,sf]{caption2}
\setcaptionmargin{4mm}

{\theorembodyfont{\sffamily} 
{\theorembodyfont{\slshape}

\textheight 240mm 
\textwidth 175 mm 
\topmargin -18mm 
  \oddsidemargin -5mm 

\begin{document}

\title{\bf Energy and angular momentum radiated for non head-on binary black hole collisions}

\author{Osvaldo M. Moreschi${}^{1}$\thanks{Member of CONICET.}, \,
        Alejandro Perez${}^{2}$
        and
        Luis Lehner${}^{3}$\thanks{CITA National Fellow and PIMS Post-Doctoral Fellow.}\\
${}^{1}${\normalsize FaMAF, Universidad Nacional de Córdoba,}\\
{\normalsize Ciudad Universitaria, (5000) Córdoba, Argentina.}\\
${}^{2}${\normalsize Center for Gravitational Physics and Geometry,}\\
{\normalsize Penn State University, University Park, PA 16802, U.S.A.}\\
${}^{3}${\normalsize Department of Physics \& Astronomy,}\\
{\normalsize The University of British Columbia, Vancouver BC, V6T 1Z1, Canada.}
}

\date{September 13, 2002}
\maketitle

\begin{abstract}
We investigate the possible total radiated energy produced 
by a binary black hole system containing non-vanishing total
angular momentum. 
For the scenarios considered we
find that the total radiated energy does not exceed $1 \%$.
Additionally we explore 
the gravitational radiation field and 
the variation of angular momentum in the process.

After the formation of the final black hole, 
the model uses the Robinson-Trautman (RT) spacetimes as background.
The evolution of perturbed RT geometries is carried out numerically.
\end{abstract}

\vspace{2mm}

\section{Introduction}
The advent of powerful detectors capable of directly
measuring gravitational radiation for the first time has
motivated numerous investigations of systems likely to 
produce gravitational waves of enough intensity which are expected to
be observed with these detectors.
Among those systems, a prime candidate is that composed
by a couple of inspiraling black holes which eventually merge
into one releasing considerable amounts of energy via gravitational
radiation. Clearly, because of the strong gravitational fields involved
in such process, its complete description
requires solving Einstein equations in their
full generality which can only be done through numerical methods\cite{luis}.
Several groups are combining efforts to numerically model such system\cite{texas,
cornell,golmnew,uiuc,psu,pitt} and although
significant progress has been (and is being) made in this
 direction (see for instance\cite{luis}),
an accurate and robust implementation is still missing.
In the mean time, valuable insight can be gained through approximate 
models of the system. These models serve both as means to gain a better understanding
of the problem and also to provide information
that can serve as tests for the numerical simulations, where the lack of
known solutions renders the problem of accuracy assessment more involved.

 Traditional models are obtained through
perturbative approximations where an expansion with respect to some
appropriately chosen parameter provides a reduced and manageable system
of equations. Naturally, these approaches provide accurate answers
only when the perturbative parameter remains small.
For instance, the traditional post-Newtonian approximation (see for instance\cite{postnewt,will}), 
can be safely used when the black holes are far enough 
and the relative
speeds involved are much smaller than 1. However, as the holes come closer, the
gravitational fields and relative speeds involved become very large and it is clear that
this approximation will fail to give sensible answers. Resumation techniques\cite{bala}
and/or effective one-body expansions\cite{damour} have been proposed to extend
the regime of validity of this approach. However, even when the ambiguities proper of
these options can be satisfactorily addressed, they will inevitably break in the late
stages of the merger, which for the case of black holes lies in the maximum sensitivity
window of earth-based detectors\cite{thorne,grischuk}.

Another approximation, known as the {\it close limit
approximation} (CLA) (see for instance\cite{pricepullin,Pullin:1999rg}) 
assumes the black holes have already merged
and the perturbation parameter can be considered to be how non-Schwarzschild
(or non-Kerr) the hole is. Since no-hair theorems imply the final
fate of the formed black hole should be of the Kerr-type, this approach
can safely be used to describe such epoch (note however, that in certain
situations, the CLA has been able to produce valuable results at rather
earlier times when the holes had just merged; however, it is not clear
that this will hold in generic cases).

A different approximate model can be obtained by considering Robinson-Trautman 
spacetimes\cite{Robinson62} which contain purely outgoing
gravitational radiation and decay to leave a Schwarzschild-like
horizon\cite{Tod89,chrusciel,robjef}. These spacetimes are thus natural candidates to 
study systems settling to a single black hole. 
The idea here would be
to use them as background
spacetimes for a perturbation treatment\footnote{Note that this approach
would therefore have a background spacetime which radiates as opposed to
considering perturbations off a stationary spacetime.}. For instance, a perturbation
approach can be constructed where the perturbation parameter is the
gravitational radiation produced  by the system\cite{OM}. This indeed 
appears as an enticing suggestion since the
system is not expected to produce total radiation outputs in excess of 
few percent of the initial mass. 

In the past, this idea was pursued to produce estimates of
the total gravitational radiation\cite{Moreschi96,Moreschi99}
obtaining excellent agreement with numerical relativity results for the head-on
collision, i.e. zero angular momentum case\cite{anninos}. These studies did not allow for the spacetime
to have total angular momentum and also the physical situations considered
produced small enough radiation
output that the equations governing
the dynamic of the perturbations were solved in the
asymptotic late-time linear regime\cite{Moreschi96,Moreschi99}. 

In order to consider more interesting scenarios, perturbations allowing for angular momentum 
were introduced in \cite{Moreschi01b} thus paving the way for studies of more generic situations.
In the present work, we perform such studies by numerically solving the dynamical
equations introduced in \cite{Moreschi01b} in their full generality. Our
studies are aimed towards obtaining a deeper understanding of the system 
and also with the hope of producing information that can be used
as a check of full numerical relativistic simulations. 

This work is organized as follows, in section \ref{sec:general} we briefly review the
perturbative treatment of Robinson-Trautman (RT)
spacetimes that allows for angular momentum.
Section \ref{sec:fixing} describes the physical parameters and how to make the
matching between the two eras that appear in our model.
The numerical implementation to solve the equations 
that govern the evolution of the perturbations is described in Section \ref{sec:numerical}. 
The numerical results are presented in Section \ref{sec:results}
where the total energy, the gravitational radiation field and the evolution
of angular momentum are shown.
We conclude with some final comments in Section \ref{sec:final}.

\section{General angular momentum perturbations of RT spacetimes}\label{sec:general}

In \cite{Moreschi01b}, generic perturbations of RT spacetimes, 
which include angular
momentum, have been presented. 

It is convenient to express the geometry in terms of a null tetrad
$(\ell ^{a},$ $m^{a},$ $\overline{m}^{a},$ $n^{a})$ where:
\begin{equation}\label{eq:produc}
g_{ab}\;\ell ^{a}\;n^{b}=-g_{ab}\;m^{a}\;\overline{m}^{b}=1
\end{equation}
with  all other possible scalar products being zero; then, the metric can be
expressed by
\begin{equation}
g_{ab}=\ell _{a}\;n_{b}+n_{a}\;\ell _{b}-m_{a}\;\overline{m}_{b}-\overline{m}%
_{a}\;m_{b}.
\end{equation}
In terms of the coordinate system $(x^{0},x^{1},x^{2},x^{3})=\left( u,r,%
(\zeta +\overline{\zeta}),\frac{1}{i}(\zeta -\overline{\zeta})\right)$,
where $u$ is a null coordinate and $r$ is an affine parameter along
the geodesic integral lines of the vector $\ell^a$,
one can express the null tetrad in terms of its components
by the relations:
\begin{equation}
\ell _{a}=\left( du\right) _{a} 
\end{equation}
\begin{equation}
\ell ^{a}=\left( \frac{\partial }{\partial \,r}\right) ^{a} 
\end{equation}
\begin{equation}
m^{a}=\xi ^{i}\left( \frac{\partial }{\partial x^{i}}\right) ^{a} 
\end{equation}
\begin{equation}
\overline{m}^{a}=\overline{\xi}^{i}\left( \frac{\partial }{\partial x^{i}}\right) ^{a} 
\end{equation}
\begin{equation}\label{eq:vecn}
n^{a}=
\,U\,\left( \frac{\partial }{\partial \,r}\right) ^{a}+X^{i}\,\left( 
\frac{\partial }{\partial \,x^{i}}\right) ^{a} 
\end{equation}
with $i=0,2,3$ and $\zeta =\frac{1}{2}\left( x^{2}+i\;x^{3}\right) $; and
where the components $\xi^{i}$, $U$ and $X^{i}$ are given by the 
following expressions:
\begin{equation}
\begin{split}
\xi ^{0} & =0, \\
\xi ^{2} & =\frac{\xi _{0}^{2}}{r} + \lambda \overline{\xi}_{0}^{2} 
\left(- \frac{\sigma_0}{r^{2}}
  + \frac{1}{r}\frac{\partial^2 W_0}{\partial r^2}
  - \frac{2}{r^{2}} \frac{\partial W_0}{\partial r}
\right)
, \\
\xi ^{3} & =\frac{\xi _{0}^{3}}{r} + \lambda \overline{\xi}_{0}^{3} 
\left(- \frac{\sigma_0}{r^{2}}
  + \frac{1}{r}\frac{\partial^2 W_0}{\partial r^2}
  - \frac{2}{r^{2}} \frac{\partial W_0}{\partial r}
\right)
\end{split}
\end{equation}
\begin{equation}
\xi _{0}^{2}=\sqrt{2}P_{0}\;V,\quad \xi _{0}^{3}=-i\xi _{0}^{2} ;
\end{equation}

\begin{equation}
U=rU_{00}+U_{0}+\frac{U_{1}}{r}+\frac{U_{2}}{r^{2}} +\Delta U_{3} ,
\end{equation}
\begin{equation}
\begin{split}
U_{00} & =\frac{\dot{V}}{V}, \\
U_{0} & =-\frac{1}{2}K_{V}, \\
U_{1} & =-\frac{1}{2}\left(\Psi _{2}^{0}+\overline{\Psi}_{2}^{0}\right) ,\\ 
U_{2} & =\frac{\lambda}{6} \left( \eth_{V_{RT}}\overline{\Psi}_{1}^{0}
             +\overline{\eth }_{V_{RT}}\Psi _{1}^{0}\right) , \\
\Delta U_3 & = - \frac{\lambda}{r^2} \left( \bar\eth_{V_{RT}}^2 W_0 
  + \eth_{V_{RT}}^2 \bar W_0 \right);
\end{split}
\end{equation}
\begin{equation}
\begin{split}
X^{0}  &=  1, \\
X^{2}  &=  \lambda \xi _{0}^{2}
          \left( -  \frac{\overline{\tau}_{0}}{r^{2}}
                 + \frac{2 \overline{\Psi}_{1}^{0}}{3r^{3}}
                 + \frac{2}{r^2}\frac{\partial \eth_{V_{RT}} \overline W_0}
                                     {\partial r}
                 - \frac{4}{r^3}\eth_{V_{RT}} \overline W_0
          \right) \\
          & \quad + {\tt c.c} , \\
X^{3}  &=  \lambda \xi _{0}^{3}
          \left( -  \frac{\overline{\tau}_{0}}{r^{2}}
                 + \frac{2 \overline{\Psi}_{1}^{0}}{3r^{3}}
                 + \frac{2}{r^2}\frac{\partial \eth_{V_{RT}} \overline W_0}
                                     {\partial r}
                 - \frac{4}{r^3}\eth_{V_{RT}} \overline W_0
          \right) \\
           & \quad + {\tt c.c} ;
\end{split}
\end{equation}
where
\begin{equation}
\tau_0 = \overline\eth_{V_{RT}}\sigma_0,
\end{equation}
${\tt c.c}$ means complex conjugate, 
\begin{equation}
K_{V}=\frac{2}{V}\overline{\eth }_{V}\eth _{V}V-\frac{2}{V^{2}}\eth _{V}V~%
\overline{\eth }_{V}V+V^{2};
\end{equation}
and where in these equations we are explicitly denoting the first order terms by
introducing the first order parameter $\lambda$ dependency, and
where $\eth_V$ is the edth operator\cite{Newman66,Goldberg67,Penrose84}, in the GHP notation\cite{GHP}, 
of the sphere with metric
\begin{equation}\label{eq:desphere}
dS^{2}=\frac{1}{P^{2}}\;d\zeta \;d\overline{\zeta} ,
\end{equation}
where $P=V(u,\zeta ,\overline{\zeta})P_{0}(\zeta ,\overline{\zeta})$, 
and $P_{0}$ is the value of $P$ for the unit sphere.

The scalar $V$ is given by
\begin{equation}
  \label{eq:vpertur}
  V=V_{RT} + \lambda \, V_\lambda ;
\end{equation}
where  $V_{RT}$ is the RT scalar satisfying the Robinson-Trautman equation
\begin{equation}
-3\, M_{0}\, \dot{V}_{RT}=V_{RT}^{4}~\eth ^{2}\overline{\eth }^{2}~V_{RT}
-V_{RT}^{3}~\eth ^{2}V_{RT}~\overline{\eth }^{2}V_{RT} ,  \label{eq:rt}
\end{equation}
$V_\lambda$ is the linear perturbation scalar and $\eth$ 
is the edth operator of the unit sphere.
One can then
express $K_V$ by
\begin{equation}
K_{V}=K_{V_{RT}}+\lambda ~K_{V_{\lambda }},
\end{equation}
where
\begin{multline}
K_{V_{RT}}=\frac{2}{V_{RT}}\overline{\eth }_{V_{RT}}\eth _{V_{RT}}V_{RT} \\
-\frac{2}{V_{RT}^{2}}\eth _{V_{RT}}V_{RT}~\overline{\eth }_{V_{RT}}V_{RT}
+V_{RT}^{2},
\end{multline}
and
\begin{multline}
K_{V_\lambda}=\frac{2}{V_{RT}}\overline{\eth }_{V_{RT}}\eth _{V_{RT}}V_\lambda
-\frac{2}{V_{RT}^{2}}\eth _{V_{RT}}V_\lambda~\overline{\eth }_{V_{RT}}V_{RT} \\
-\frac{2}{V_{RT}^{2}}\eth _{V_{RT}}V_{RT}~\overline{\eth }_{V_{RT}}V_\lambda 
+\frac{2V_\lambda}{V_{RT}^{3}}\eth _{V_{RT}}V_{RT}
         ~\overline{\eth }_{V_{RT}}V_{RT} \\
+ V_\lambda \, V_{RT} + \frac{V_\lambda}{V_{RT}} \, K_{V_{RT}}
.
\end{multline}

In the above equations we have $\sigma_0=\sigma_0(u,\zeta,\bar\zeta)$,
$W_0=W_0(u,r,\zeta,\bar\zeta)$, $\Psi_1^0=\Psi_1^0(u,\zeta,\bar\zeta)$
and
\begin{multline}
\Psi_2^0=\Psi_2^0(u,\zeta,\bar\zeta) \\
= -\biggl( M_0+ \lambda \bigl(M_1(u,\zeta,\bar\zeta) + i \, \mu(u,\zeta,\bar\zeta)
\bigr) \biggr);
\end{multline}
where $\mu$ is related to $\sigma_0$ by
\begin{equation}\label{eq:mu}
\mu = \frac{1}{2i}\left(\eth_{V_{RT}}^2 \overline\sigma_0 -
                    \overline\eth_{V_{RT}}^2 \sigma_0 \right)
.
\end{equation}

In order to study the intrinsic fields at future null infinity (scri),
it is convenient to consider the leading order behavior of $W_0$, namely
\begin{equation}
W_0 =  \frac{\Psi_0^0}{4!r} + W_1 ;
\end{equation}
where $\Psi_0^0=\Psi_0^0(u,\zeta.\bar\zeta)$ and $W_1 =O(1/r^2)$.
It is interesting to note that the component 
$\Psi_0 =  \frac{\partial^4 W_0}{\partial r^4} $
of the Weyl tensor is given in this case\cite{Moreschi01b} by
\begin{equation}
\Psi_0=  \frac{\Psi_0^0}{r^5} + \frac{\partial^4 W_1}{\partial r^4},
\end{equation}
where the second term is of order $O(1/r^6)$.

The remaining equations at scri are:
\begin{equation}\label{eq:psi0u}
\dot\Psi_0^0 = 3\frac{\dot V_{RT}}{V_{RT}}\Psi_0^0 + \eth_{V_{RT}}\Psi_1^0
 -3M_0 \,\sigma_0 ,
\end{equation}

\begin{equation}\label{eq:psi1u}
\dot \Psi_1^0=
 3\frac{\dot V_{RT}}{V_{RT}} \Psi_1^0 - \eth_{V_{RT}} (M_1 + i\mu)
 - \overline\eth_{V_{RT}}\left(K_{V_{RT}}\right)\sigma_0 
\end{equation}
and
\begin{multline}\label{eq:vla}
-6M_0 \frac{\dot{V}_{\lambda }}{V_{RT}} =
\;\overline{\eth }_{V_{RT}}\eth _{V_{RT}}~K_{V_{\lambda }}
- 6\frac{\dot{V}_{RT}}{V_{RT}}\left(3M \frac{V_\lambda}{V_{RT}} 
- M_{1}\right) \\
- 2\dot{M}_{1}  
-   \frac{\dot{V}_{RT}}{V_{RT}} \overline\eth_{V_{RT}}^2 \sigma_0 
+ 2 \frac{\dot{V}_{RT}}{V_{RT}^2} \overline\eth_{V_{RT}} V_{RT} 
                                  \overline\eth_{V_{RT}}\sigma_0 \\
- \frac{2}{V_{RT}}  \overline\eth_{V_{RT}} \dot V_{RT} 
                                  \overline\eth_{V_{RT}}\sigma_0 
+  \overline\eth_{V_{RT}}^2 \dot{\sigma}^0 \\
-   \frac{\dot{V}_{RT}}{V_{RT}} \eth_{V_{RT}}^2 \overline\sigma_0 
+ 2 \frac{\dot{V}_{RT}}{V_{RT}^2} \eth_{V_{RT}} V_{RT} 
                                  \eth_{V_{RT}}\overline\sigma_0 \\
- \frac{2}{V_{RT}}  \eth_{V_{RT}} \dot V_{RT} 
                                  \eth_{V_{RT}}\overline\sigma_0
+ \eth_{V_{RT}}^2 \dot{\overline\sigma}^0 
.
\end{multline}

The objective is to use the solutions of the previous equations
to model a black hole that
just formed after the non head-on collision of a previous
binary system. 
In this model the idea is to only make use of the information
of the individual masses and the total angular momentum. 
With all this in
mind we have to chose the appropriate gauge and free functions.

In reference \cite{Moreschi01b} it was discussed the gauge freedom of
these spacetimes. In order not to introduce extra
structure, we will chose the free functions and gauges that make
\begin{equation}
M_1=0 ;
\end{equation}
and
\begin{equation}
\sigma_0=0 .
\end{equation}
Other choices will force extra
structure on the model beyond masses and angular momentum; as mentioned
above.

With this choice the previous equations become
\begin{equation}\label{eq:psi0u2}
\dot\Psi_0^0 = 3\frac{\dot V_{RT}}{V_{RT}}\Psi_0^0 + \eth_{V_{RT}}\Psi_1^0
,
\end{equation}

\begin{equation}\label{eq:psi1u2}
\dot \Psi_1^0=
 3\frac{\dot V_{RT}}{V_{RT}} \Psi_1^0 
\end{equation}
and
\begin{equation}\label{eq:vla2}
-6M_0 \,\dot{V}_{\lambda } =
\;V_{RT}\overline{\eth }_{V_{RT}}\eth _{V_{RT}}~K_{V_{\lambda }}
- 18 M_0 \dot{V}_{RT}\left( \frac{V_\lambda}{V_{RT}} \right)
.
\end{equation}

It is interesting to note that since the RT equation can also be expressed as
\begin{equation}\label{eq:rt2}
-6M_0 \frac{\dot{V}_{RT}}{V_{RT}}=\;\overline{\eth }_{V_{RT}}\eth
_{V_{RT}}\,K_{V_{RT}}, 
\end{equation}
equation (\ref{eq:vla2}) can be written in the following way
\begin{multline}
\label{eq:vla3}
-6M_0 \,\dot{V}_{\lambda } =
\;V_{RT}\overline{\eth }_{V_{RT}}\eth _{V_{RT}}~K_{V_{\lambda }} \\
+
3 V_\lambda
\overline{\eth }_{V_{RT}}\eth_{V_{RT}}\,K_{V_{RT}} .
\end{multline}

Given initial conditions for the functions $V_{RT}$, $\Psi_0^0$, $\Psi_1^0$
and $V_\lambda$ equations (\ref{eq:rt}), (\ref{eq:psi0u2}), (\ref{eq:psi1u2})
and (\ref{eq:vla2}) provide for the respective evolutions.

The radiation content of the spacetime is easily described by the time derivative
of the Bondi shear. Let $\sigma_B^0$ denote the Bondi shear and $u_B$ the Bondi 
time; then for any section of scri $u=\text{\sf constant}$ one can express
the radiation content from the relation\cite{Dain96}
\begin{equation}\label{eq:sigmBadot}
\frac{\partial \sigma_B^0}{\partial u_B}= \frac{\eth^2 V}{V} ;
\end{equation}
therefore to first order one has
\begin{equation}\label{eq:sigmBadot2}
\frac{\partial \sigma_B^0}{\partial u_B}= 
\frac{\eth^2 V_{RT}}{V_{RT}}
+\lambda \left(\frac{\eth^2 V_\lambda}{V_{RT}}
        -\frac{V_\lambda\eth^2 V_{RT}}{V_{RT}^2}
\right).
\end{equation}

The radiation flux at the retarded time $u$ is given by
\begin{equation}\label{eq:fb}
F_B(u) = \frac{1}{4\pi}\int_{S_u} 
\frac{\partial \sigma_B^0}{\partial u_B}
\frac{\partial \bar\sigma_B^0}{\partial u_B} \, dS^2 .
\end{equation}

Let us recall that in this case the 
timelike component of the Bondi momentum at the section $u=\tt constant$ 
can be calculated from
\begin{equation}\label{eq:Bondimass}
P^0_B=\frac{1}{4\pi} \int \frac{M_0}{V^3} dS^2 .
\end{equation}

At the retarded time $u$, we say that $V$ represents a quadrupole excitation 
along the $y$ axis if 
\begin{equation}
V= 1 + A Y_{2,0}(\zeta',\bar\zeta');
\end{equation}
where $A$ is the amplitude of the excitation, $Y_{2,0}$ is an spherical
harmonic, and $(\zeta',\bar\zeta')$ are the coordinates of the sphere
where the pole is along the $y$ axis. Then in terms of the usual 
spherical harmonics with coordinates $(\zeta,\bar\zeta)$, one has
\begin{equation}
V= 1 
- \frac{A}{2} 
\left(
\sqrt{\frac{3}{2}} Y_{2,-2}(\zeta,\bar\zeta)
+ Y_{2,0}(\zeta,\bar\zeta) 
+\sqrt{\frac{3}{2}} Y_{2,2}(\zeta,\bar\zeta)
\right)
.
\end{equation}

\section{Fixing the parameters and the Newtonian matching}\label{sec:fixing}

The physical system can be characterized by two stages. 
During the first stage two black holes are moving towards each other
with some orbital angular momentum. At some moment they collide
and form a single black hole; which settles down in the asymptotic
future to a stationary Kerr geometry.

During the first stage of evolution we describe the gravitational
radiation with the quadrupole formula, where the dynamics is worked
out from the Newtonian framework. The whole motion is contained in a
plane; therefore two variables are sufficient to describe the orbits.
The two integrals of motion, namely energy $E$ and angular momentum $j$, 
allow to solve the Newtonian system.

The initial data is assumed to be given in the $(x,y)$ plane; with
zero total momentum and such that the orbital angular momentum
is along the positive $\hat z$ direction

The initial velocities can be thought to have components  along the $x$
axis only. Let us call $R_0$ the initial impact parameter, at some initial relative 
distance $r_0$. In this way, using the relative velocity $v_0$, one
has
\begin{equation}
E=\frac{\mu}{2}v_0^2 - \frac{G m_1 m_2}{r_0},
\end{equation}
\begin{equation}
j=\mu R_0 v_0;
\end{equation}
where $\mu=\frac{m_1 m_2}{m_1 + m_2}$ is the reduced mass.


For this kind of motion the quadrupole formula predicts that the 
power of gravitational radiation is given by
\begin{equation}\label{eq:pq}
F_Q(r)=\frac{8}{15}(G m_1 m_2)^2 \left[\frac{2}{\mu r^4} 
\left(E + \frac{G m_1 m_2}{r} \right) + \frac{11 j^2}{\mu^2 r^6} \right];
\end{equation}
which generalizes the analog equation appearing in ref. \cite{Moreschi99}.

In order to continue the dynamical description after the collision of the 
two black holes, we need to have a merging condition in the Newtonian
framework. In references \cite{Moreschi96} and \cite{Moreschi99} we have
succeeded in estimating the total energy radiated in the head on black
hole collision using the following criteria. When a black hole, which
mass at infinity is $m_{i1}$, is brought to a distance $r_{12}$ of another
black hole of asymptotic mass $m_{i2}$, its physical mass, for the 
stationary situation, is 
changed\cite{Brill63} to
\begin{equation}
m1 = m_{i1} + \frac{m_{i1} m_{i2}}{2 r_{12}};
\end{equation}
and similarly for the other mass.

Applying the arguments of ref. \cite{Moreschi99} one concludes that
the merging condition should be taken when the separation distance has
the value
\begin{equation}
r_c = 2 \left( m1 + m2 \right).
\end{equation}

It is convenient to introduce the relative mass parameter $\alpha$ 
and the reference mass $m$, so that $m_{i1}=m$ and $m_{i2}=\alpha m$.
Then the merging condition\cite{Moreschi99} gives
\begin{equation}
r_c = (m_{i1} + m_{i2})
\left[1+\sqrt{1+\frac{2\alpha}{(1+\alpha)^2}} \right].
\end{equation}

Let us now consider the initial data used in ref. \cite{texas} for the
grazing collisions of black holes; namely:
$x_1=5m$, $x_2=-5m$, $y_1=m$, $y_2=-m$, $v_{x1}=-0.5$ and $v_{x2}=0.5$;
with $m=1$ and $\alpha=1$. This data corresponds to a hyperbolic 
Newtonian trajectory; with initial separation distance 
$r=10.198$, and critical merging radius $r_c = 4.449$.

Our strategy is to follow closely the model used in \cite{Moreschi96,Moreschi99}; but
we also want to compare our work with \cite{texas}.
However, the initial data of \cite{texas} involves two black holes with
half the speed of light each; therefore in adapting this initial data
to our model we must take into account relativistic effects.
In references \cite{Moreschi96} and \cite{Moreschi99}
there was no need for these concerns because the initial data
was not relativistic.

Since the initial velocities are relativistic, 
the relative initial velocity $v_0$
is calculated from the expression
\begin{equation}\label{eq:sumvel}
v_0 = \frac{v_{x2} -  v_{x1}}{1 - v_{x1} v_{x2}}=0.8;
\end{equation}
where we are using geometric units in which the velocity of light
and the gravitational constant have the unit value.

The  energy radiated during the falling phase, calculated from (\ref{eq:pq})
is $E_N = 0.00346 (m_1+m_2)$.

In order to match the Newtonian stage with the black hole RT perturbed model
we need also to set the total initial mass and initial angular momentum
for the RT stage. 

In previous work we have matched the Newtonian mass $M_{New}=m_{i1}+m_{i2}$
to the initial mass $M$; this already takes into account the field relativistic
first order correction in terms of its physical masses $m1$ and $m2$, since,
recalling equation (21) of ref. \cite{Moreschi99}, 
the initial mass would be
$m1+m2 - \frac{m_{i1}m_{i2}}{r_{12}}$. 
The system discussed in ref.  \cite{Moreschi99} had zero initial velocity, and
therefore there was no need to take into account any other effect. 
Instead in our case we should take into account speed relativistic
corrections.

Then, since the initial
data is relativistic, the initial mass and angular momentum
are calculated from
\begin{equation}\label{eq:relativisticmass}
M= \frac{m_{i1}}{\sqrt{1-v_1^2}} + \frac{m_{i2}}{\sqrt{1-v_2^2}} =2.309
\end{equation}
and
\begin{equation}
J=-y_1 \frac{m_{i1} v_{x1}}{\sqrt{1-v_{x1}^2}} -y_2 \frac{m_{i2} v_{x2}}{\sqrt{1-v_{x2}^2}}
= 1.155 .
\end{equation}
It should be emphasized that the relation between the initial RT stage
mass and angular momentum $M$ and $J$ with the
Newtonian mass and angular
momentum $M_{New}$ and $j$ is just the Lorentzian factor 
$\gamma=\frac{1}{\sqrt{1-v_{x1}^2}}=1.1547$.
In the last section we will comment on the incidence of this factor 
on our results.

Let us observe that 
the relation between the relativistic angular momentum and total mass is
$\frac{J}{M^2}=0.217$.

Since the angular momentum is small, it can be treated as a perturbation.
At the moment of the collapse, we can consider a quadrupole excitation along
the $y$ axis with amplitude $A$, as described above. Let us take 
$A= A_m + A_j$; where $A_j$ is the contribution coming from the
appearance of the angular momentum $j$. The matching condition is given
by the equation (See equations (\ref{eq:sigmBadot}), (\ref{eq:fb}) and (\ref{eq:pq}).)
\begin{equation}
F_B = F_Q(r_c) ;
\end{equation}
from which one obtains in this case
\begin{equation}
A_m = 0.057
\end{equation}
and
\begin{equation}
A_j = 0.028
.
\end{equation}
Therefore, to be explicit, the initial data for $V_{RT}$ and $V_\lambda$ are
\begin{multline}
V_{RT} = 
1
 - \frac{A_m}{2} \\
\times \left(
\sqrt{\frac{3}{2}} Y_{2,-2}(\zeta,\bar\zeta)
+ Y_{2,0}(\zeta,\bar\zeta)
+\sqrt{\frac{3}{2}} Y_{2,2}(\zeta,\bar\zeta)
\right)
;
\end{multline}
and
\begin{multline}
\lambda \, V_\lambda = 
 - \frac{A_j}{2} \\
\times \left(
\sqrt{\frac{3}{2}} Y_{2,-2}(\zeta,\bar\zeta)
+ Y_{2,0}(\zeta,\bar\zeta)
+\sqrt{\frac{3}{2}} Y_{2,2}(\zeta,\bar\zeta)
\right)
.
\end{multline}


The constant $M_0$ is determined from  (\ref{eq:Bondimass}),
and the condition that initially the mass is given by (\ref{eq:relativisticmass});
which sets $M_0 = 2.302 $.

The orbital angular momentum is taken into account in the initial data for
$\Psi_1^0$.
It is convenient to express this initial data in terms
of the auxiliary field $g$, given by
\begin{equation}
\Psi_1^0 = i \frac{\eth_{V_{RT}} g}{V_{RT}^2} = i \frac{\eth g}{V_{RT}}.
\end{equation}
Let us note that then
\[
\eth_{V_{RT}} \Psi_1^0 = \eth^2 g
\]
so that in the stationary case one has $\eth^2 g =0$ and $\dot g=0$.

We take $g=\bar g$ and
\begin{equation}
g =g_0 \, Y_{1,0}(\zeta,\bar\zeta) ;
\end{equation}
where $g_0$ is related to the angular momentum in the $z$ direction by
\begin{equation}
g_0 = -\sqrt{12 \pi} J = -4.094 .
\end{equation}

In reference \cite{Bruegmann99} it was also considered a similar case of a binary 
system with orbital angular momentum. Their initial data was:
$x_1=0$, $x_2=0$, $y_1=m$, $y_2=-m$, $v_{x1}=-0.8$ and $v_{x2}=0.894$;
with $m=1.5$ and $\alpha=\frac{2}{3}$. From the Newtonian point of view
this data corresponds to an elliptic motion; but with maximum and minimum
radius that are smaller than the corresponding critical merging
radius. For this reason, we can not compare this case with our model.

\section{Numerical Implementation}\label{sec:numerical}
\label{numerical}
Accurate numerical evolution of a fourth order parabolic equation, such
as (\ref{eq:rt}), by means of an explicit finite difference scheme is
a challenge because the $CFL$\cite{gustafsson} condition requires that the time step
$\Delta u$ scale as the fourth power of the spatial grid size.
Nevertheless, we constructed a set of algorithms to solve these equations
using second
order accurate finite difference approximations  (following \cite{eth}). The numerical treatment 
of the {\it eth} operator has been thoroughly described
in \cite{eth}. This work presented a clean way to deal with
derivative operators on the sphere by covering it with two coordinate
patches and dealing with spin weighted quantities. Thus, it is ideally
suited for our present purposes. The numerical grid on each patch is defined by
$\xi_{ij} = q_{i} + i p_{j}$ where ${q_i, p_i} = -1 - 2 \Delta_A + (i-1) \Delta_A$
(with $\Delta_A=2/(N_A-5)$). The angular derivatives are discretized
by centered second order finite difference approximations and information
between patches is obtained through fourth order accurate interpolations. (For
a detailed description of this approach see\cite{eth}).

The integration in time is  based upon a three time level 
Adams-Bashford~\cite{gustafsson} scheme with predictor
($\tilde{\cal F}$) given by
\begin{equation}
  {\tilde {\cal F}}(u+\Delta u)={\cal F}(u)
      +{\Delta u \over 2} \partial_u[3 {\cal F}(u)-{\cal F}(u-\Delta u)],
\end{equation}
and corrector
\begin{multline}
   {\cal F}(u+\Delta u)={\cal F}(u)
 +{\Delta u \over 2} \partial_u[ {\cal F}(u)+{\tilde {\cal F}}(u+\Delta u)] \\
  +O(\Delta u^3).
\end{multline}
Where ${\cal F}$ stands for $V_{RT}$ or $V_{\lambda}$ and
the $\partial_u$ terms are to right hand sides of equations (\ref{eq:vla}).
Additionally, we implemented the iterative
Cranck-Nicholson algorithm\cite{teukolsky,alcubierre} and observed
that the results obtained with both implementations agree.
Since the evolution equation for $\Psi^{0}_{0}$ is linear, its numerical
integration is straightforwardly done by centered second order differences
at the level $(u+\Delta u/2)$.

The second order convergence of numerical solutions was confirmed
in the perturbative regime using solutions of the linearized equation and
second order self-convergence of the solutions was confirmed in 
the non-linear regime.

\section{Results}\label{sec:results}

{\bf  \large Variations of total energy}

The total Bondi energy-momentum vector at any RT retarded time $u$ 
can be calculated from the expression
\begin{equation} \label{P}
P^{\tt a}=-\frac{1}{4\pi} \int_S l^{\tt a}(\zeta, \bar \zeta) 
\left( \Psi_{B2}^0 + \sigma_B \dot{\bar \sigma}_B \right)
\,  dS^2, 
\end{equation}
where $S$ is the section determined by $u={\tt constant}$,
\begin{equation} \label{la}
 (l^{\tt a}) =\left( 1,\frac{\zeta +\bar \zeta }
{1+\zeta \bar{\zeta }},\frac{\zeta -\bar{\zeta }}{i(1+\zeta 
\bar{\zeta )}},\frac{\zeta \bar{\zeta }-1}{1+\zeta \bar{\zeta 
}}\right) ;
\end{equation}
and 
the subscript $B$ is used to emphasize that the quantities are
evaluated with respect to a Bondi frame.
The mass $M$ at  this section $S$ is then given by
\begin{equation} \label{M}
M=\sqrt{P^a P_a},
\end{equation}
where the indices are raised and lowered by the Lorentzian 
flat metric $\eta_{ab}$ at scri\cite{Moreschi86}.

Let us note that the relations between the Bondi quantities and the
intrinsic ones are
\begin{equation}
  \label{eq:psi2bondipsi2}
   \Psi_{B2}^{0} = \frac{ \Psi_{2}^{0}}{V^3}
\end{equation}
and
\begin{equation}
  \label{eq:sigmabondisigma}
  \sigma_B = \frac{\sigma}{V} ;
\end{equation}
therefore in our gauge one has $\sigma_B = 0$, at each RT section.

The gravitational energy radiation flux is calculated from the Bondi
time derivative of the supermomentum $\Psi$\cite{Moreschi88}; namely
\begin{equation}
  \label{eq:psiubondi}
  \frac{\partial \Psi}{\partial u_B} =  \frac{\partial \sigma_B}{\partial u_B} \,
                                        \frac{\partial {\bar\sigma}_B}{\partial u_B} .
\end{equation}
If one instead considers the time change with respect to the RT time,
it is convenient to have in mind that for any function $f$ one has
\begin{equation}
  \label{eq:ubondiurt}
   \frac{\partial f}{\partial u_B} =  \frac{1}{V} \frac{\partial f}{\partial u} .
\end{equation}

The so called news function $ \frac{\partial \sigma_B}{\partial u_B}$ can be
expressed in terms of the perturbed RT fields by
\begin{equation}
  \label{eq:news}
   \frac{\partial \sigma_B}{\partial u_B} = 
     \frac{\eth^2 V}{V} +  \frac{1}{V^2}\left(\dot\sigma -  \frac{\dot V}{V}\sigma \right).
\end{equation}

To calculate the total energy radiated, one could then numerically evaluate 
the gravitational energy radiation flux of equation (\ref{eq:psiubondi}) at
different times and sum along all the elapsed time.
However, it is more accurate to numerically evaluate the initial mass and
subtract the final mass. This is due to the fact that the RT spacetime is 
known to converge asymptotically to the Schwarzschild one; more 
specifically, one knows that $\lim_{u \rightarrow \infty} V_{RT} = 1$;
and similarly one can see that  $\lim_{u \to \infty} V_\lambda = 0$.

Using this procedure, and a resolution of 
$n=32$ points for
half a meridian of the sphere (approximately $N=1600$ 
points for
the whole sphere),
the energy radiated $E_{RT}$ during the RT stage is 
found to be $E_{RT}= 0.0034 M_0$.

Then, since in units of $M_0$, the energy radiated in the first stage is
 $E_N = 0.0030 M_0$, the total energy radiated in the whole process
is $E = 0.0064 M_0$.

{\bf \large Gravitational radiation field}

\begin{figure}[h]
\centering
\includegraphics[clip,width=0.5\textwidth]{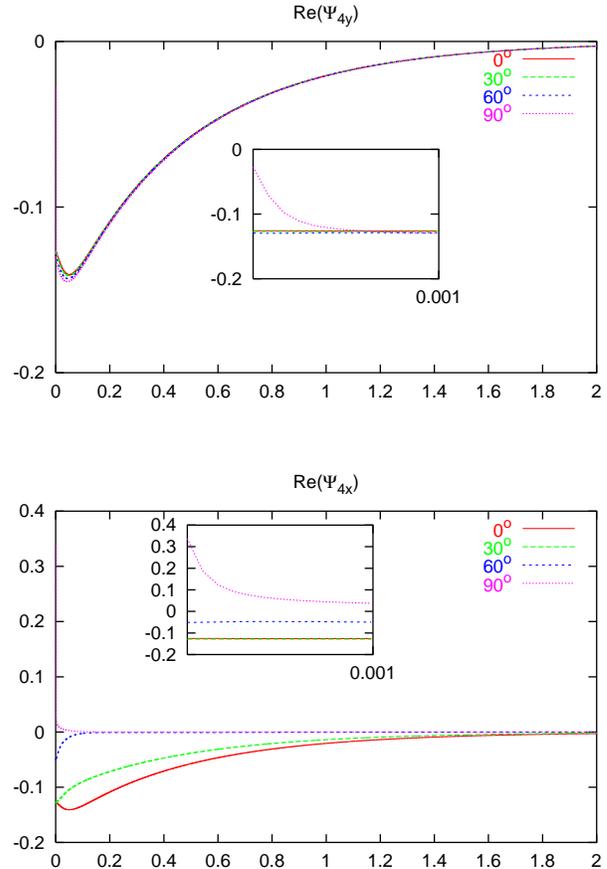}
\caption{Evolution of $\Psi_4$ as a function of time. 
The subscript $x$ and $y$ refer to the $x$ and $y$ axis
of the stereographic coordinate $\zeta=x+i y$ of the
northern hemisphere.
Each of the four curves refer to detectors at $0^\circ$,
$30^\circ$, $60^\circ$ and $90^\circ$; measured from the
north pole of the sphere.
The insets show the detail of the
rapid time variations at the beginning.
}
\label{fig:psi4}
\end{figure}

The numerical calculation of the evolution of the gravitational radiation
field $\Psi_4$, is shown in figure \ref{fig:psi4}.

It can be seen that although the total energy radiated is rather small,
the amplitude of the gravitational radiation field can be large.
In other words, this model describes a noticeable burst.

This is interesting since  $\Psi_4$ is precisely what 
gravitational wave detectors will measure. 

{\bf \large Variations of angular momentum}

When dealing with the notion of angular momentum one is faced with the
fact that there are several inequivalent definitions of angular momentum;
which are not tightly related with the notion of intrinsic angular momentum,
with the exception of \cite{Moreschi01c}. An appropriate definition
of intrinsic angular momentum involves the selection of unique
sections\cite{Moreschi01c}
of future null infinity where the quantity is to be calculated.
Since in our case we are taking the angular momentum as a perturbation
parameter of the RT geometries, it is not essential to consider these
refinements in our model. And also, since the RT spacetimes provide
with a geometric unique family of sections of future null infinity,
namely the sections $u=\tt constant$, it is natural to use them
to calculate the angular momentum.

Then, instead of describing the variation of the intrinsic
angular momentum we describe the variation of the RT-angular momentum vector
given by
\begin{equation}
  \label{eq:ji}
  J^k =  \Re \mathrm{e} \left[ \int_{S_{RT}}  \, \frac{i}{4\pi} \bar\eth \ell^k 
    \Psi_{B1}^{0}  \;dS_B^{2} \right] ;
\end{equation}
where $S_{RT}$ are the sections determined by $u=\tt constant$, 
$k=1,2,3$, so that $\ell^k$ are the spacelike components of $\ell^{\tt a}$;
and where the Bondi component $\Psi_{B1}^{0}$, of the Weyl tensor, is related to the RT
Weyl component $\Psi_{1}^{0}$ by
\begin{equation}
  \label{eq:psi1bondipsi1}
   \Psi_{B1}^{0} = \frac{ \Psi_{1}^{0}}{V^3}.
\end{equation}

\begin{figure}[hbtp]
\centering
\includegraphics[clip,width=0.35\textwidth,angle=-90]{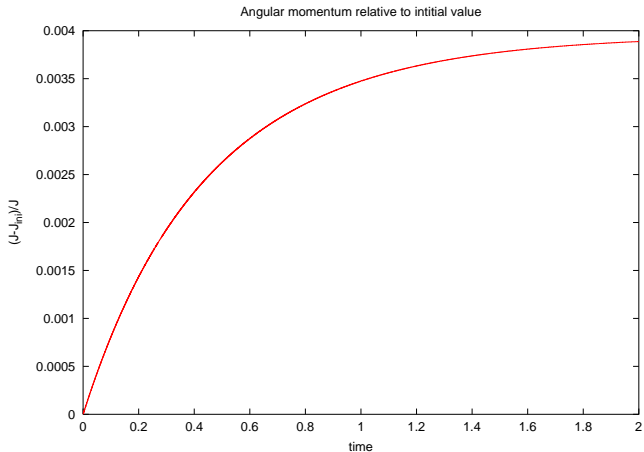}
\caption{Evolution of the relative variation of angular momentum $J$
as a function of time.
}
\label{fig:angrel}
\end{figure}

Figure \ref{fig:angrel} shows a very small and smooth variation of the angular
momentum; which is more related to the time variation of the RT geometry
than to the radiation of angular momentum; as can be seen from
the nature of equation (\ref{eq:psi1u2}).

\section{Final comments}\label{sec:final}
The estimate of the total gravitational energy radiated in the non head-on 
collision of two black holes is less than one percent according to our model.
This is on the low end value of existing numerical results (which are still 
being refined) and comparable with those obtained with the close limit 
approximation\cite{pullin}.
It is however worth mentioning  that our calculations predict an amplitude of the 
gravitational waves that could be important from the observational point of view. Additionally,
it predicts a rather narrow time duration of the `burst'. Since the frequency at which this
happens; its amplitude and the time duration can be used as preliminary information for
constructing data analysis `filters'\cite{anderson,creighton}, knowledge of these
is of importance while templates from full numerical simulations are 
not
available. 
We will carry out a detailed study of these in the future.

At first sight there is too large a difference between the numerical results
in \cite{texas} and the estimates obtained here. However, one should keep in mind
several points which (by themselves) can account for this difference. Comments are in place
for both the results in \cite{texas} and those presented here:

First,
the results in \cite{texas} were obtained under a rather coarse resolution and 
the final numbers need to be refined (there are intense efforts worldwide in this
direction). Additionally, the obtained value of radiated energy come from 
comparisons of initial ADM mass and apparent horizon masses. Masses obtained from
apparent horizon calculations, in dynamical regimes, are only an approximation of
the mass the black hole and hence results obtained through this method can have a
significant systematic error. Furthermore
it is important to remember that the initial data of  \cite{texas} is 
given at a spacelike hypersurface  of
the spacetime; while, by the nature of the RT spacetime, our initial data is given on
a characteristic surface. This implies an essential difference since in the
spacelike hypersurface initial data one has incoming and outgoing radiation
contributing to the total energy radiated;
while in the RT characteristic problem that we solve,
we do not consider any incoming initial radiation in our calculations. 
So, for the relativistic regime, one expects the time slice initial data 
to radiate more energy than the characteristic case.

As far as the present model is concerned: in previous works we have obtained 
very good agreements between our estimates
and the mature numerical calculations of exact geometries; as one can check in the 
family of systems depicted in figures 3 and 4 of reference \cite{Moreschi99}.
The fact that the model represented so well such a variety of initial data
and different mass ratios, motivated us to apply the same techniques to 
the case in which angular momentum was involved. However, since there are just
two (still being refined) numerical calculations available to compare with, it is difficult
to obtain much information at this stage.

As we have mentioned before our aim is to follow as close as possible the 
model used in \cite{Moreschi96}
and \cite{Moreschi99}; 
however the fact that the initial conditions considered are relativistic
posse several questions.
For example, in matching the masses of the two stages we have used
equation (\ref{eq:relativisticmass}); so one might ask 
how would
the results change if the relativistic $\gamma$ factor was not used?
This would certainly decrease the initial $M$ and $M_0$ by about 13\%;
but $E_N$ and $E_{RT}$ would not change, so that the ration between the
total energy radiated versus the changed $M_0$ would increase by the factor
$\gamma$, that is about 15\%.

Similarly, had we not used the relativistic correction on the angular momentum
$J$, its initial value would have decreased by about 13\%; but the relative variation
shown in figure \ref{fig:angrel} would not have changed.

In any case, we think that by applying our model to this relativistic initial
data, we are pushing the model to the boundary of its validity. Comparisons with
future numerical simulations will shed light on this and indicate how far this model
can be pushed.

The radiation of angular momentum seems to be negligible with these initial conditions.
In order to consider higher values of the initial angular momentum, we
would need to deal with other background geometries, as for example
twisting algebraically special spacetimes. 
Regarding the smooth monotonic variation of it one can infer that, 
for this small initial angular momentum data, 
its behavior is driven by the exponential asymptotic behavior
of the RT background geometry.
There are not complicated initial variations in $\Psi_1^0$, that
for example do appear in $\Psi_4^0$, as seen in figure \ref{fig:psi4}.

When describing a concrete physical situation with these spacetimes, one is supposed
to choose the gauge and fix the free functions in order to make the best 
representation of the system. It is somehow striking that the choice of the frame in
first order has physical significance, and it is not pure gauge as one is accustomed
in the studies of linearized gravity around Minkowski spacetime.

\section*{Acknowledgments}

We acknowledge support from CONICET,
SeCyT-UNC, NSF Grants No. PHY9900791 and PHY0090091, and Eberly Research Funds 
of Penn State. Some computations were performed in the VN cluster (vn.physics.ubc.ca)
supported by the Canadian Foundation for Innovation.

We are grateful to an anonymous Referee for suggesting several improvements and
pointing out some typos.




\end{document}